%% file: Anomalies_submit_cond.tex
\documentclass[prl,twocolumn,letterpaper]{revtex4}%
\usepackage{amsmath}
\usepackage{graphicx}
\usepackage{graphics}
\usepackage{amsfonts}
\usepackage{amssymb}
\usepackage{color}  
\begin{document}
\newcommand{\vett}[1]{ \mathbf{#1} }
\title{Conductance and density of states anomalies  as resonances of energy bands in disordered coupled chains}
\author{L. Alloatti$^1$ and G. Grosso$^2$}
\affiliation{$^1$Max-Planck Institut f\"ur Metallforschung, Heisenbergstr. 3, D-70569 Stuttgart, Germany and Institut f\"ur Theoretische und Angewandte Physik, Universit\"at Stuttgart, Pfaffenwaldring 57, D-70569 Stuttgart, Germany\\
$^2$NEST-INFM and Dipartimento di Fisica "E. Fermi", Universit\`a di Pisa, Via F. Buonarroti 2, I-56127 Pisa, Italy}

\begin{abstract}
We show that off-diagonal nearest neighbor disorder in  quasi-one-dimensional single particle tight-binding coupled chains leads to anomalies in the density of states and  in the mean conductance,  that can be interpreted as due to specific resonances of the  band structure of the perfect system underlying the disordered one. We demonstrate that this phenomenology may appear not necessarily at the energy $E=0$ as reported so far in the literature and we show that also the even-odd chain number effect on the mean conductance is ruled by the same resonances. 
For different cases we provide a series expansion for the resonant contribution to the mean conductance. These expansions allow to make predictions well verified numerically.
\end{abstract}


\maketitle
The existence of a divergent density of states at the center of the spectrum ($E=0$) of a one-dimensional phonon model with off-diagonal disorder goes back to the pioneer paper of Dyson in 1953~\cite{Dyso53}. The effect of this kind of disorder on the divergence of the localization length at the energy $E=0$ was then evidenced ~\cite{Theo75,Egga78}, in contrast with the general wisdom that in strictly one dimensional systems with uncorrelated diagonal or off-diagonal disorder all the states are localized~\cite{Borl63,Souk81}.

Recently further interest has grown on off-diagonal disorder in quasi-one-dimensional systems composed of $N$ coupled chains. In this case it has been shown that for odd $N$ the localization length and the density of states diverge at the center of the band, at zero energy, while they remain finite for even $N$\cite{insiemeII}.

It has been claimed that the origin of delocalization of the states at  $E=0$, is the existence of an additional  (chiral) sublattice symmetry  present in system with off-diagonal disorder only~\cite{InsiemeIII}.

To our knowledge  only the point $E=0$ has been investigated in the literature. In this paper we look for the possibility of occurrence of the anomalies described  above at other energy points.  We show that such anomalies can arise in every point of the energy spectrum, provided  a well defined condition on the energy bands of the perfect underlying chain system is fulfilled.

We start considering an infinitely extended periodic quantum wire composed by $N$ interacting chains; diagonal and off-diagonal disorder is then introduced in a strand of length $\mathcal{L}$ of the wire (the lattice constant is taken as unit of length). 

The  Schr\"odinger equation for the multichain system considered is:
\begin{eqnarray}
U_n \psi_n + T_{n-1} \psi_{n-1}+T_n^{\dagger} \psi_{n+1}=E \psi_n,{} & \forall n,
\end{eqnarray}
where $U_n$ is the matrix connecting the sites belonging to the n-th column  and  $T_n$ is the matrix specifying the hopping amplitudes from the column $n$ to the next one. The order of these matrices is equal to  the number of coupled chains, $N$. The index $n\in {\mathbb Z}$ labels the columns along the wire and the $\psi_n$ are  ${N}$-dimensional vectors.

In order to introduce the disorder we write the matrices $U_n$ and $T_n$ in the following form:
\begin{eqnarray}
U_n=U+W^{U}_{n}, & \quad T_n=T+W^T_{n},
\label{distinzione}
\end{eqnarray}
where $U$ and $T$ are the matrices, independent of $n$, defining the periodic chain system and $W^U_{n}$ and $W^T_{n}$ are randomly distributed matrices. We take $W^U_{n}\neq 0$ only for $n=1,...,\mathcal{L}$  and $W^T_{n}\neq 0$ for $n=1,...,\mathcal{L}-1$. The distinction \eqref{distinzione} defines the ordered part and the perturbation part (disorder) of the Hamiltonian, namely $H=H_0+W$. All the elements of the random matrices have the form 
\begin{equation}
(W^{U(T)}_n)_{i,j}=w^{U(T)}_{i,j}\chi_{i,j} e^{i\alpha^{U(T)}_{i,j} \chi^{'}_{i,j}}
\end{equation}
where all the $\chi$ and $\chi^{'}$ are independent and uniformly distributed real random variables, with zero mean and second moment equal to one. The terms $w^{U(T)}_{i,j}$ and $\alpha^{U(T)}_{i,j}$ do not depend on the column index $n$ and regulate the modulus and the phase of the disorder.

Let us consider first the periodic chain system described by the Hamiltonian $H_0$. Let
 $Z(\theta_B)$ be the matrix diagonalizing the hermitian matrix:
\begin{equation}
Q=Te^{-i \theta_{B}}+U+T^{\dagger} e^{i \theta_{B}},
\label{condizione_di_banda}
\end{equation}
where $\theta_B\in[-\pi,\pi]$ is the "Bloch angle", and the eigenvalues $E_i(\theta_B) $ form the band structure of the unperturbed system. 
 The states 
\begin{equation}
\psi^{(j)}_n \equiv Z^{(j)} e^{in\theta_B}, 
\label{stato_di_Bloch}
\end{equation}
where  $Z^{(j)}$ denotes the $j$-th column of the matrix  $Z(\theta_B)$, 
form a complete set of eigenstates of  $H_0$.

The Landauer conductance $g(E)$ of the disordered wire can be calculated by means of the M\o{}ller operator~\cite{scattering}
\begin{equation}
\Omega_{+}(E)=1+\frac{1}{ (H-E+i\epsilon) }W=1+G^{R}(E)W,
\end{equation}
where $G^R(E)$ is  the retarded Green's function of the disordered  system. $\Omega_{+}(E)$ is an $\infty \times \infty $ matrix, but its non-trivial part has finite dimension.

If we indicate with $P_B$ the projector on the region of length $(\mathcal{L}+2)$ containing the disordered region plus the two unperturbed columns located at $n=0$ and at $n=\mathcal{L}+1$ and by $P_b$ the projector on the column of index $\mathcal{L}+1$, the transmission amplitude at the energy $E$ from the energy channel  $\eta$ on the left, to the energy channel  $\nu$ on the right  on the disordered region is given by
\begin{equation}
A^{\tau}(E,\eta \rightarrow \nu)=\langle E,\nu| P_{b}\Omega_{+}P_B|E,\eta\rangle,
\label{ampiezza}
\end{equation}
where $|E,\eta(\nu)\rangle $ are the Bloch states of energy $E$ belonging to the $\eta(\nu)$-th band (see Eq. \eqref{stato_di_Bloch}) and with positive group velocity.
The conductance is calculated by means of the following equation:
\begin{equation}
g(E)=2e \sum_{\eta,\nu}D(E,\eta) \vert A^{\tau}(\eta \rightarrow \nu)\vert ^2 v(E,\nu),
\label{conduttanza}
\end{equation}
where $v(E,\nu)=\frac{1}{\hbar} \frac{\partial E(\theta,\nu)}{\partial \theta}$ and $D(E,\eta)=\frac{1}{2\pi}\left( \frac{\partial E(\theta,\eta)}{\partial \theta} \right)^{-1}$ are respectively the group velocity on the band $\nu$ and the density of states per column of the underlying perfect wire at the energy $E$ on the  band $\eta$;  $e$ is the electron charge. 
We choose the units in such a way that the conductance $g(E)$ of the perfect wire assumes integer values.
 
 If the matrix $Z$ is independent of the Bloch angle $\theta_B$ \cite{nota1}, and at the energy $E$ {\it all}  the bands of the perfect wire are present,  it can be shown that the expression  \eqref{ampiezza} assumes the following particularly simple form \cite{seguito}:   
\begin{eqnarray}
A^{\tau}(\eta \rightarrow \nu) \! & \! = \! & 
\langle \left(
 \begin{array}{c}
0                   \\
0                   \\
\vdots  \\
\hat{e}_{\nu}             \\
\end{array}\right)
\!|\frac{1}{ 1-\mathbb{F} \widetilde{\mathbb{G}} \widetilde{\mathbb{W}}}|\!
\left( \!
\begin{array}{c}
\hat{e}_{\eta}  \\
\hat{e}_{\eta}e^{i \theta_{\eta}^+ } \\
\vdots  \\
\hat{e}_{\eta}e^{i (\mathcal{L}+1) \theta_{\eta}^+ } \\
\end{array}
\! \right)\rangle . 
\label{omega_coniugata}
\end{eqnarray} 
The meaning of the symbols entering in Eq. \eqref{omega_coniugata} is specified below:
\begin{itemize}
\item
$\hat{e}_{\nu}$ is the standard vector of dimension $N$ having one at the $\nu$-th position and zero elsewhere.
\item 
$\mathbb{\widetilde{W}}=\mathbb{Z}^{\dagger}(P_B W P_B) \mathbb{Z}$. 
\item $\mathbb{Z}$, $\mathbb{\widetilde{G}}$ are block-diagonal matrices with blocks $Z$ and $\widetilde{G} \equiv Z^{\dagger}GZ$, respectively. $G$ is the \emph{free} Green's function projected on the left and on the right of the same column. Under the hypothesis just assumed,  $\widetilde{G}$ is diagonal with values $\widetilde{G}_{\eta,\eta}=-i2\pi D(E,\eta)$.
\item
\begin{minipage}{5cm}
\begin{equation}
\mathbb{F}=\left(
\begin{array}{cccccc}
\mathbb{I}  & F_{1}^-   & \cdots    & F_{\mathcal{L}+1}^-      \\
F_{1}^+  & \mathbb{I}     & \cdots    & F_{\mathcal{L}}^-  \\
\vdots  & \vdots  & \ddots    & \vdots      \\
F_{\mathcal{L}+1}^+  & F_{\mathcal{L}}^+   & \cdots    & \mathbb{I}      \\
\end{array}
\right),
\nonumber
\end{equation}
\end{minipage}

where $F_{n}^{\pm}$ are the $N \times N$ diagonal matrices defined by $(F_{n}^{\pm})_{\eta,\eta}=e^{\pm in\theta_{\eta}^{\pm}}$. Here $\theta_{\eta}^{\pm}$ are the Bloch angles at which the $\eta$-th band has energy E and positive (+) or  negative (-)  group velocity.
\end{itemize}
Eq.~\eqref{omega_coniugata} is the main scattering theory tool for the evaluation of disorder effects on the conductance of a periodic system composed by an arbitrary number $N$ of chains.  To exemplify its use, we first treat the case of a strictly one-dimensional system;  the case of two coupled chains will be discussed successively.

\begin{figure}[htbp]
\begin{center}
\input{1.pstex_t}
\end{center}
\caption{First diagram contributing to the mean squared conductance in the regime of resonance ($\theta^+-\theta^-=\pi$).}
\label{primo_diagramma}
\end{figure}
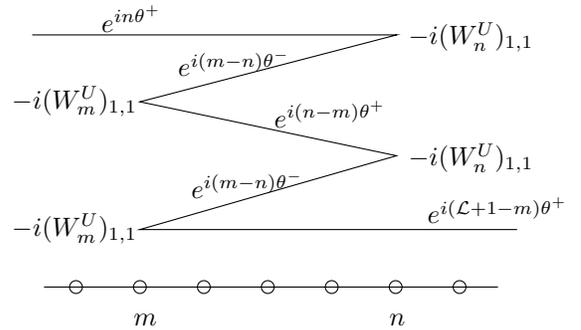

For the strictly one-dimensional model ($\eta$=$\nu$=1) with pure $diagonal$ disorder in the region $\mathcal{L}$, the expansion of the fraction in Eq. \eqref{omega_coniugata} gives:
\begin{tabular}{l}
$A^{\tau}=e^{i (\mathcal{L}+1) n\theta^+}$ \\
$+f \ \sum_n \ \  \ e^{i(\mathcal{L}+1-n)\theta^+}(\chi_{n,n})e^{in\theta^+}$ \\
$+f^2 \sum_{m,n}e^{i(\mathcal{L}+1-m)\theta^+}(\chi_{m,m})e^{i(m-n)\theta^{\pm}}(\chi_{n,n})e^{in\theta^+}$\\
$+\cdots$\\
\end{tabular}
where $f\equiv-i2\pi D(E,1)w^U_{1,1}$ and the sums are over all the sites within the disordered region; moreover, $\theta^{\pm}$ is equal to $\theta^+$ if $m>n$ and $\theta^-$ if $m<n$.
Using Eq.~\eqref{conduttanza} the average conductance is  
\begin{equation}
\langle g(E)\rangle =\langle|A^{\tau}|^2\rangle.
\end{equation}
Every term contributing to the average conductance can be represented by a Feynman diagram. We have found that  it is convenient to classify all these diagrams into two categories: those for which their phase depends from the particular sites where the scatterings took place and the others. 
At the lowest order (fourth), there is just one non-vanishing diagram belonging to the latter category (see Fig. \ref{primo_diagramma}).
This diagram, in fact, is proportional to the squares of the disorder parameters $(\chi_{n,n})^2$ and $(\chi_{m,m})^2$ so that, taking the average value it does not vanish; moreover,   moving either the position $n$ or the position $m$ by one or more lattice parameters in either directions, it acquires a total phase change equal to $e^{i2(\theta^+-\theta^-)}$. The mechanism by which the resonance appears as a sharp peak is that only for $(\theta^+-\theta^-)=q\pi$, with $q$ integer, {\it all the amplitudes of such diagrams sum coherently}. We name the occurrence of this condition as "$\pi$- coupling of the energy bands".
 
It is possible to see that, as far as the condition $(\theta^+ - \theta^-)=\pi$ is satisfied, not only the fourth order of the average conductance acquires an additional term, but also every even-order term of the expansion  in the disorder strength, so determining the magnification of the average conductance. We define the sum of all these diagrams as the "resonant contribution to the average conductance" $\langle g\rangle _{res}$.

For one-dimensional systems the resonance occurs at $E=0$. With $T_{1,1}=-\frac{1}{2}$ (which implies, at $E=0$, $G_{1,1}=-i$ and $\theta^+ = -\theta^-=\frac{\pi}{2}$) and for diagonal disorder of intensity $w^U_{1,1}=w_d$ the sum of all the diagrams up to the twelfth order gives:
\begin{eqnarray}
\langle g(E=0)\rangle _{res}=\frac{2}{2!}w_d^4\mathcal{L}^2-\frac{42}{3!}w_d^6\mathcal{L}^3+\frac{888}{4!}w_d^8\mathcal{L}^4 \nonumber\\
-\frac{24336}{5!}w_d^{10}\mathcal{L}^5+\frac{1521631}{6!}w_d^{12}\mathcal{L}^6+{}\cdots
\label{sviluppo_wd}
\end{eqnarray}
With similar manipulations  for \emph{real} off-diagonal disorder of intensity $w^T_{1,1}=w_f$ we obtain:
\begin{eqnarray}
\langle g(E=0)\rangle _{res}= \frac{32}{2!}w_f^4\mathcal{L}^2-\frac{1664}{3!}w_f^6\mathcal{L}^3+\frac{112640}{4!}w_f^8\mathcal{L}^4\nonumber\\
-\frac{10698752}{5!}w_f^{10}\mathcal{L}^5+\frac{1392902144}{6!}w_f^{12}\mathcal{L}^6+\cdots 
\label{sviluppo_wf}
\end{eqnarray}
Interestingly, we have noticed numerically that for any given value of $w_d$ there exists a value of $w_f$ such that the resulting conductances $g(E)$ have almost the same off-resonance shape, while the height of the peak for off-diagonal disorder is much bigger than the other. In addition, we note that for off-diagonal disorder all the diagrams contributing to a fixed order in the expansion have the same sign, while for diagonal disorder there is a mutual cancellation of diagrams.

 We have extended the previous approach to  quantum wires composed by  $N$ different chains  all with equal  nearest-neighbor interactions and containing a disordered region always of the same kind and length. We have observed a strong dependence of the average conductance at $E=0$ from the number $N$ of chains. In particular, for $N$ odd and a wide range of real off-diagonal disorder \cite{seguito}, $\langle g(E=0;N)\rangle$ is considerably greater than for the even case $N+1$. By comparing  this result with the conductance evaluated at slightly different energies (where the resonance is destroyed), we can conclude  that this even-odd effect of the conductance is just caused by the resonant terms discussed above.

In the case of more than one coupled quantum chains, the rule for calculating the diagrams which are relevant is very similar to the one depicted  for the one dimensional case. The only difference consists in the fact that now there are more angles which can propagate and additional care has to be taken in selecting the paths contributing to the resonance.  We give in Eq. \eqref{trace} below the fourth order expansion of  the resonant contribution to the conductance  of the two coupled chains with two resonating bands:
\begin{eqnarray}
&\langle g \rangle _{res}=\frac{\mathcal{L}^2}{2!}\langle \mbox{Tr}
    &(     P_1 M_m P_1 M_n P_2 M_m P_2 M_n P_1 \nonumber\\ 
&     &   +  P_1 M_m P_2 M_n P_1 M_m P_2 M_n P_1 \nonumber\\
&     &   +  P_1 M_m P_2 M_n P_2 M_m P_1 M_n P_1 \label{trace}\\
&     &   +  P_2 M_m P_2 M_n P_1 M_m P_1 M_n P_2 \nonumber\\
&     &   +  P_2 M_m P_1 M_n P_2 M_m P_1 M_n P_2 \nonumber\\
&     &   +  P_2 M_m P_1 M_n P_1 M_m P_2 M_n P_2)\rangle+c.c.  \nonumber
\end{eqnarray}
where $n$ and $m$ are two arbitrarily chosen different column indexes and:
\begin{eqnarray} 
M_i=Z^\dagger GW_{U,i}Z +F^- Z^\dagger GW_{T,i}^{\dagger}Z+Z^\dagger GW_{T,i}ZF^+; \nonumber
\end{eqnarray}
\begin{eqnarray}
P_1 = \left(
\begin{array}{cc}
1  & 0 \\
0 & 0  \\
\end{array}
\right); &
P_2 = \left(
\begin{array}{cc}
0  & 0 \\
0 & 1  \\
\end{array}
\right); &
F^\pm = \left(
\begin{array}{cc}
e^{\pm i \theta_1^\pm}  & 0 \\
0 & e^{\pm i \theta_2^\pm}  \\
\end{array}
\right). \nonumber
\end{eqnarray}
The result of the previous trace is in general quite cumbersome, so we report it only for the case where the Bloch angles are $\theta_1^{+}=-\theta_1^{-}=\frac{\pi}{3}$, and  $\theta_2^{+}=-\theta_2^{-}=\frac{2\pi}{3}$, and the matrices $Z$ and $G$ are given by:
\begin{eqnarray}
Z \equiv  \left(
\begin{array}{cc}
\cos\alpha  & -\sin\alpha \\
\sin\alpha & \cos\alpha  \\
\end{array}
\right); &
\widetilde{G} = \left(
\begin{array}{cc}
-\frac{i}{\sqrt{3} }  & 0 \\
0 & -\frac{i}{\sqrt{3} }  \\
\end{array}
\right).
\label{matrice_zeta}
\end{eqnarray}
It is easy to verify that for $\alpha=\frac{\pi}{4}$ we recover the conductance at $E=0$ of two coupled chains with first neighbor hoppings all equal to $-1$.

\begin{figure}[ht]
\begin{center}
\includegraphics[width=3in]{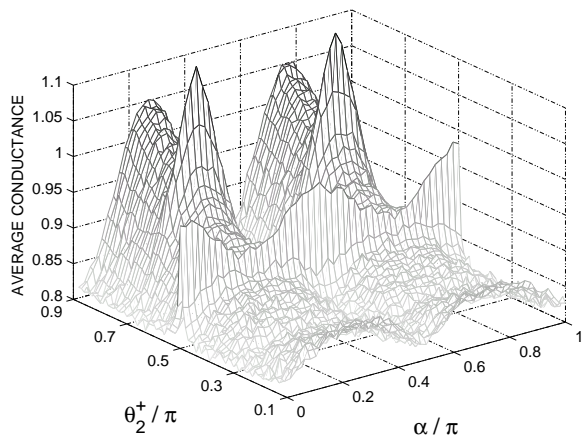} 
\end{center}
\caption{Average conductance for two coupled chains as function of the Bloch angle $\theta_2^{+}=-\theta_2^{-}$ and of the $\alpha$ angle defining the matrix $Z$ in \eqref{matrice_zeta}. The angle $\theta_1^{+}=-\theta_1^{-}$ are fixed to be $\pm \frac{\pi}{3}$. The peaks appearing for $\theta_2^{+}=\frac{2\pi}{3}$ are due to the resonance contribution and, as function of $\alpha$, it follows qualitatively the $\sin^4\alpha$ predicted by  \eqref{quarto_ordine_C2_analitico1}. A gray-scale has been used to emphasize the peaks. }%
\label{grafico_fasi}
\end{figure}

\begin{figure}[ht]
\begin{center}
\includegraphics[width=3in]{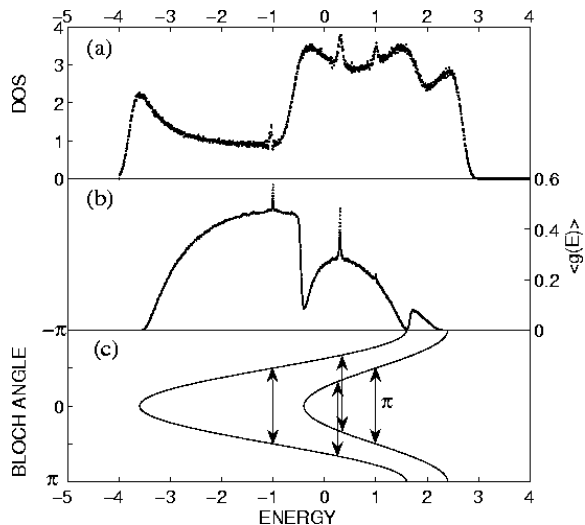}
\end{center}
\caption{Average DOS in arbitrary units (a), average conductance (b),  and band structure(c) of the two coupled chains defined by \eqref{matrici_speciali}. The density of states corresponds to the disordered region (device) isolated from the  leads. There are three energies at which  $\pi$-coupling occurs. In correspondence of these energies, the average conductance and the density of states are peaked.}
\label{speciale}
\end{figure}

Introducing the disorder, of equal intensity, only on the hoppings corresponding to the matrix elements $(W^T_n)_{1,1}$ and $(W^T_n)_{2,2}$, we obtain:
\begin{eqnarray}
\langle g(\alpha)\rangle _{res}=\frac{11}{6}\left( \Re \langle (W^T_{1,1})^2\rangle \right)^2 \mathcal{L}^2\sin^4\alpha+\mathcal{O}(w^6\mathcal{L}^3).
\label{quarto_ordine_C2_analitico1}
\end{eqnarray}
This $\sin ^4 \alpha$ behavior is in good qualitative agreement with the numerical data reported in Fig. \ref{grafico_fasi}.
The most interesting cases are those for which the $\pi$-coupling between the bands may occur for energies different from the center of the energy spectrum or only between a subgroup of all the bands. In this cases, in fact, the (chiral) sublattice symmetry claimed in the references  \cite{InsiemeIII} is broken. An example of this occurs for
\begin{eqnarray}
U = \left(
\begin{array}{cc}
0  & -1 \\
-1 & 0  \\
\end{array}
\right); &
T = \left(
\begin{array}{cc}
-1  & -0.3 \\
-0.3 & -1  \\
\end{array}
\right).
\label{matrici_speciali}
\end{eqnarray}
In Fig.~\ref{speciale} (a) we report the density of states and in Fig.~\ref{speciale} (b) the average conductance for the two-chains system with off-diagonal disorder. The energy band structure of the perfect underlying system is  reported in Fig. \ref{speciale} (c).
As a further example let us focus on the peak at $E=1$ where  all the bands are present but only one resonates.
In this case, an equation similar to  Eq. \ref{trace}  gives:
\begin{eqnarray}
\langle g\rangle _{res}\propto \Re \left(\langle (W^T_{1,1})^2\rangle+\langle (W^T_{2,2})^2\rangle -\langle (W^U_{1,2})^2\rangle  \right)^2 \mathcal{L}^2+\cdots \nonumber
\label{quarto_ordine_C2_analitico2}
\end{eqnarray}
The latter expression shows clearly that these two kinds of off-diagonal disorder (on the $T$ and on the $U$  matrices) compete with each other, and this is consistent with our numerical simulations \cite{seguito}. 

From the evaluation of the spectrum of eigenvalues for several configurations we have obtained the average DOS depicted in Fig. \ref{speciale} (a). This figure evidences a direct connection between the peaks in the conductance and in the density of states. As far as we know, such a connection has been demonstrated for strictly one dimensional systems only \cite{Thou72}.

Finally we justify why we kept in the development of the mean conductance only terms of the form $w^{2p}\mathcal{L}^q$ with $p=q$ and no smaller values of $q$. The reason is that the average conductance $\langle g(E;w,\mathcal{L})\rangle $ converges at a fixed non-trivial value for every energy when $\mathcal{L} \rightarrow \infty$ with $w^2\mathcal{L}$ constant ($w$ represents the global strength of disorder). In practice we have verified that this convergence is very fast \cite{seguito}.
 
We would like to thank A. Cresti for having provided a very fast code, based on the Keldysh formalism \cite{Grosso}, used as a check of our code and as a source of large statistics of the conductance.

\end{document}

%% file: 1.pstex_t
\begin{picture}(0,0)%
\includegraphics{1.pstex}%
\end{picture}%
\setlength{\unitlength}{3522sp}%
\begingroup\makeatletter\ifx\SetFigFont\undefined%
\gdef\SetFigFont#1#2#3#4#5{%
  \reset@font\fontsize{#1}{#2pt}%
  \fontfamily{#3}\fontseries{#4}\fontshape{#5}%
  \selectfont}%
\fi\endgroup%
\begin{picture}(4297,2230)(1,-1856)
\put(631,254){\makebox(0,0)[lb]{\smash{{\SetFigFont{10}{12.0}{\familydefault}{\mddefault}{\updefault}{\color[rgb]{0,0,0}$e^{in\theta^+}$}%
}}}}
\put(1846,-421){\makebox(0,0)[lb]{\smash{{\SetFigFont{10}{12.0}{\familydefault}{\mddefault}{\updefault}{\color[rgb]{0,0,0}$e^{i(n-m)\theta^+}$}%
}}}}
\put(1261,-916){\makebox(0,0)[lb]{\smash{{\SetFigFont{10}{12.0}{\familydefault}{\mddefault}{\updefault}{\color[rgb]{0,0,0}$e^{i(m-n)\theta^-}$}%
}}}}
\put(2926,-1096){\makebox(0,0)[lb]{\smash{{\SetFigFont{10}{12.0}{\familydefault}{\mddefault}{\updefault}{\color[rgb]{0,0,0}$e^{i(\mathcal{L}+1-m)\theta^+}$}%
}}}}
\put(856,-1816){\makebox(0,0)[lb]{\smash{{\SetFigFont{10}{12.0}{\familydefault}{\mddefault}{\updefault}{\color[rgb]{0,0,0}$m$}%
}}}}
\put(2656,-1816){\makebox(0,0)[lb]{\smash{{\SetFigFont{10}{12.0}{\familydefault}{\mddefault}{\updefault}{\color[rgb]{0,0,0}$n$}%
}}}}
\put(1171,-61){\makebox(0,0)[lb]{\smash{{\SetFigFont{10}{12.0}{\familydefault}{\mddefault}{\updefault}{\color[rgb]{0,0,0}$e^{i(m-n)\theta^-}$}%
}}}}
\put(2791,164){\makebox(0,0)[lb]{\smash{{\SetFigFont{10}{12.0}{\familydefault}{\mddefault}{\updefault}{\color[rgb]{0,0,0}$-i(W_n^U)_{1,1}$}%
}}}}
\put(  1,-286){\makebox(0,0)[lb]{\smash{{\SetFigFont{10}{12.0}{\familydefault}{\mddefault}{\updefault}{\color[rgb]{0,0,0}$-i(W_m^U)_{1,1}$}%
}}}}
\put(2791,-691){\makebox(0,0)[lb]{\smash{{\SetFigFont{10}{12.0}{\familydefault}{\mddefault}{\updefault}{\color[rgb]{0,0,0}$-i(W_n^U)_{1,1}$}%
}}}}
\put(  1,-1186){\makebox(0,0)[lb]{\smash{{\SetFigFont{10}{12.0}{\familydefault}{\mddefault}{\updefault}{\color[rgb]{0,0,0}$-i(W_m^U)_{1,1}$}%
}}}}
\end{picture}%

%% file: Anomalies_submit_cond.bbl
\begin{thebibliography}{99}
\bibitem{nota1} Necessary and sufficient condition is that the matrices $U$ and $T$ are both diagonalizable and that they have a common basis of eigenvectors. A sufficient condition is $T=T^{\dagger}$ and $[U,T]=0$. 
 In these cases the bands are cosine-like, namely $E_{\nu}(\theta)=\lambda^{(U)}_{\nu}+2|\lambda^{(T)}_{\nu}|\cos(\theta-\theta_{\nu})$, where $\lambda^{(U,T)}_{\nu}$ are the eigenvalues of the matrices $U$ and $T$ and $\theta_{\nu}=\arctan[\Im(\lambda^{(T)}_{\nu})/\Re(\lambda^{(T)}_{\nu})]$.
\end{thebibliography}
